# Insights into Cold Source MOSFETs with Sub-60 mV/decade and Negative Differential Resistance Effect

*Yiheng Yin*[†], *Zhaofu Zhang*[‡, *], *Chen Shao*[†], *John Robertson*[†,‡] *and Yuzheng Guo*[†, *]

† School of Electrical and Automation, Wuhan University, Wuhan, Hubei 430072, China

‡ Department of Engineering, University of Cambridge, Cambridge, CB2 1PZ, United Kingdom

**Abstract：** To extend the Moore's law in the 5 nm node, a large number of two dimensional (2D) materials and devices have been thoroughly researched, among which the "cold" metals 2H $MS_2$ (M = Nb, Ta) with unique band structures are expected to achieve the sub-60 mV/dec subthreshold swing (SS). The studied "cold" metal field-effect transistors (CM-FETs) based on the "cold" metals are capable to fulfill the high-performance (HP) and low-dissipation (LP) goals simultaneously, as required by the International Technology Roadmap for Semiconductors (ITRS). Moreover, gaps of "cold" metals also enable the CM-FETs to realize negative differential resistance (NDR) effect. Owing to the wide transmission path in the broken gap structure of $NbS_2/MoS_2$ heterojunction, the recording 4110 μA/μm peak current, several orders of magnitude higher than the tunneling current of the Esaki diode, is achieved by $NbS_2/MoS_2$ CM-FET. The largest peak-valley ratio (PVR) $1.1 \times 10^6$ is obtained by $TaS_2/MoS_2$ CM-FET with $V_{GS}$ = -1V at room temperature. Our results claim that the superior on-state current, SS, cut-off frequency and NDR effect can be obtained by CM-FETs simultaneously. The study of CM-FETs provides a practicable solution for state-of-the-art logic device in sub 5 nm node for both more Moore roadmap and more than Moore roadmap applications.

## 1. INTRODUCTION

The Moore's law now faces the bottleneck due to the short channel effect of Si-based metal-oxide-semiconductor field field-effect transistors (MOSFET).[1-2] Considering the prospect of integrated circuits (ICs), two main technology roadmaps have been put forward by the industry field, namely more Moore roadmap and more than Moore roadmap.[3-4] The more Moore roadmap aims to pursue the device scaling complying with the Moore's law,[3] while the more than Moore technology focuses on improving the functionality of ICs, such as combining digital electronics with devices such as radio frequency (RF), power devices and sensors.[4] In the more Moore field, unlike the obvious thickness-dependent mobility of Si and Ge channels, two dimensional (2D) materials not only have atomically thin channel thickness which is beneficial for the gate control, but also maintain promising carrier mobility.[5-8] Previous works have proven that devices using 2D materials are capable to fulfill the International Technology Roadmap for Semiconductors (ITRS) requirements[9] even with nanometer-scale gate length.[10-15] Recently, the transistor channel length has already scaled down to 5 nm node.[16] How to further decrease the power consumption and sustain the Moore's law is a task of top priority. Various novel transistors such as Fin filed-effect-transistors (FinFETs),[17] fully depleted silicon on insulator (FDSOI) transistors,[18] tunneling FET (TFETs)[19-20] and negative capacitance (NC) transistors[21-22] have been put forward. Apart from novel device configurations, looking for new materials is an alternative approach to sustain the Moore's law. Recently, cold-source FETs (CS-FETs) have been proposed to achieve sub-60 mV/dec subthreshold swing (SS), which can be realized by using materials with desired density of state (DOS) such as Dirac materials[23], appropriately doped semiconductors[24-25] and materials with gaps near the Fermi level ($\varepsilon_F$).[26] Compared with the complex heterostructure fabricated by Dirac source materials or appropriately doped semiconductors, the "cold" metal 2H $MS_2$ (M = Nb, Ta) with gaps close to the $\varepsilon_F$, equivalent to a naturally p-doped or n-doped semiconductor, would be an ideal solution to fulfill the steep slope of SS.[26] The unique band structure allows the "cold" metal MOSFET (CM-FET) to filter the transmission of high-energy carriers in the subthreshold region and reach sub-60 mV/dec SS.[23] Another cornerstone is that the "cold" metal monolayer 2H $NbS_2$ and $TaS_2$ have been successfully synthesized[27-28] and served as the injection source in the heterojunction transistors,[29] which solids the way for the research of "cold" metal heterojunctions and transistors.

In this work, we conduct a comprehensive electronic and transport calculation of "cold" metals ($NbS_2$ and $TaS_2$) and their devices with transition metal dichalcogenide (TMD) channels. The SS of CM-FETs successfully breakthrough the 60 mV/dec thermionic limit at room temperature. In terms of more Moore field (to extend the Moore's law), CM-FETs are capable to fulfill the on-state current, power consumption and cut-off frequency ($f_T$) requirements of both ITRS high performance (HP) and low dissipation (LP) goals. Apart from the favorable performance against ITRS goals, the CM-FETs with unique band structures of "cold" metals can successfully achieve the negative differential resistance (NDR) effect, which is expected to fulfill the multifunctional ICs in the more than Moore field. Owing to the wide transmission path in the broken gap characters of $MS_2/MoS_2$ heterojunctions, the peak current is several orders of magnitude higher than the typical tunneling current of Esaki diode,[30-31] whose operation principle is shown in Figure S1. The benchmarking 4110 μA/μm peak current and $1.1 \times 10^6$ peak-valley ratio (PVR) are achieved by $NbS_2/MoS_2$ and $TaS_2/MoS_2$ CM-FETs, respectively. The results claim that the superior $I_{on}$, SS, $f_T$ and NDR effect are obtained by our CM-FETs simultaneously, which provides a feasible method for the development of state-of-the-art logic devices beyond the Moore's law.

## 2. RESULTS

### 2.1 The performance of CM-FETs against the ITRS goals

In the more Moore domain, we only focus on whether the transistors can satisfy the ITRS goals and extend the Moore's law. The calculated lattice constants of monolayer 2H $NbS_2$ and $TaS_2$ are $a_1$ = 3.36 Å and $a_2$ = 3.31 Å respectively, which is in agreement with previous results.[29, 32] The band structures and DOS of 2H $NbS_2$ and $TaS_2$ are shown in Figure 1. Energy gaps occur above ($E_{CG}$) and below ($E_{VG}$) the $\varepsilon_F$ of "cold" metals. The $NbS_2$ and $TaS_2$ can thus be considered as heavily p-doped or n-doped semiconductors and used as the injection source of a transistor. Considering the DOS between the $\varepsilon_F$ and $E_{VG}$ is higher than the DOS above the $\varepsilon_F$, the "cold" metal is more suitable to serve as the source electrode in a p-type transistor. Previous work presented that $MX_2$ CM-FETs with 10 nm gate length have successfully reached the sub-60 mV/dec SS at room temperature.[26] To illustrate the mechanism of superior SS, schematic energy band diagram comparisons of the p-type conventional MOSFET and CM-FET in off-state are shown in Figure 1(c, d).

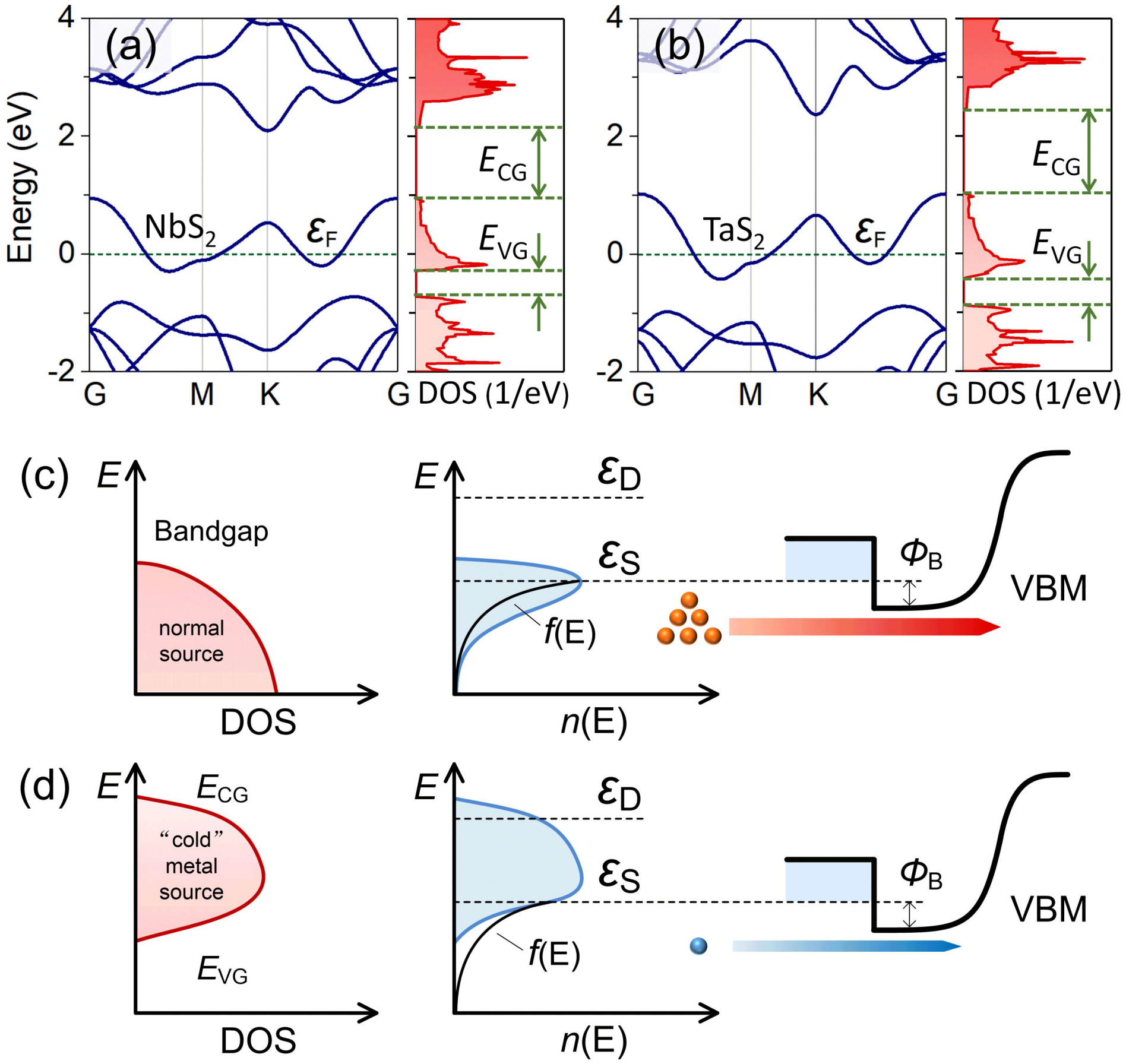

**Figure 1**. (a, b) Band structures of monolayer metallic 2H $MS_2$ (M = Nb, Ta). The green dash lines in band structures represent the Fermi level. The energy gaps in the conduction band ($E_{CG}$) and valence band ($E_{VG}$) are labelled. (c, d) The schematic density of states (DOS), carrier density ($n$(E)) and energy band diagrams of the (c) conventional MOSFET and (d) CM-FET. $\varepsilon_S$ and $\varepsilon_D$ represent the Fermi level of source and drain, respectively. $\Phi_B$ is the channel barrier height. The width of arrows represents the magnitude of currents, and the number of balls represent the carrier density for transmission. The black curves in $n$(E) represent the Fermi-Dirac distribution.

The SS is defined as

$$\mathrm{SS} = \frac{dV_{\mathrm{GS}}}{d(\lg I_{DS})} = \ln 10\left(\frac{k_{\mathrm{B}}T}{q}\right)\left(\frac{C_{\mathrm{OX}} + C_{\mathrm{S}}}{C_{\mathrm{OX}}}\right) \tag{1}$$

where $V_{GS}$ is voltage applied between gate and source electrodes. $I_{DS}$ is the drain-source current. $k_B$ and $T$ are the Boltzmann constant and temperature, respectively. $q$ is the electronic charge. $C_{OX}$ and $C_S$ are the oxide capacitance and semiconductor capacitance, respectively. The thermionic limit of 60 mV/decade is owing to the constant term $\ln 10(\frac{k_{\mathrm{B}}T}{q}) \approx 60$. Because carriers in the source have an energy distribution complying to the Fermi-Dirac distribution $f$(E).[33] The energy distribution of carriers is defined as $n$(E) = $g$(E)$f$(E), where $g$(E) is the DOS distribution as a function of energy. Based on Eq. (1), the SS is proportional with $T$. With the decreasing of $T$, the $f$(E) around the Fermi level becomes sharp, leading to the sub-60 mV/dec SS. Similarly, unlike the $n$(E) of normal source having a long thermal tail, the thermal tail of "cold" metal source below $\varepsilon_S$ is filtered by the $g$(E) around $E_{VG}$.[23] The $n$(E) of "cold" metal source, steeper than the $f$(E), decreases super-exponentially with the decreasing of energy, which allows the CM-FET to achieve the sub-60 mV/dec SS. Meanwhile, electrons localized around the $\varepsilon_S$ permit a large on-state current. The schematic energy band diagrams of the CM-FET corresponding to the on-state and the off-state are shown in Figure S2. As the transistor channel length scales down, it is of interest to investigate whether the superior SS can be maintained in the sub-5 nm node CM-FETs.

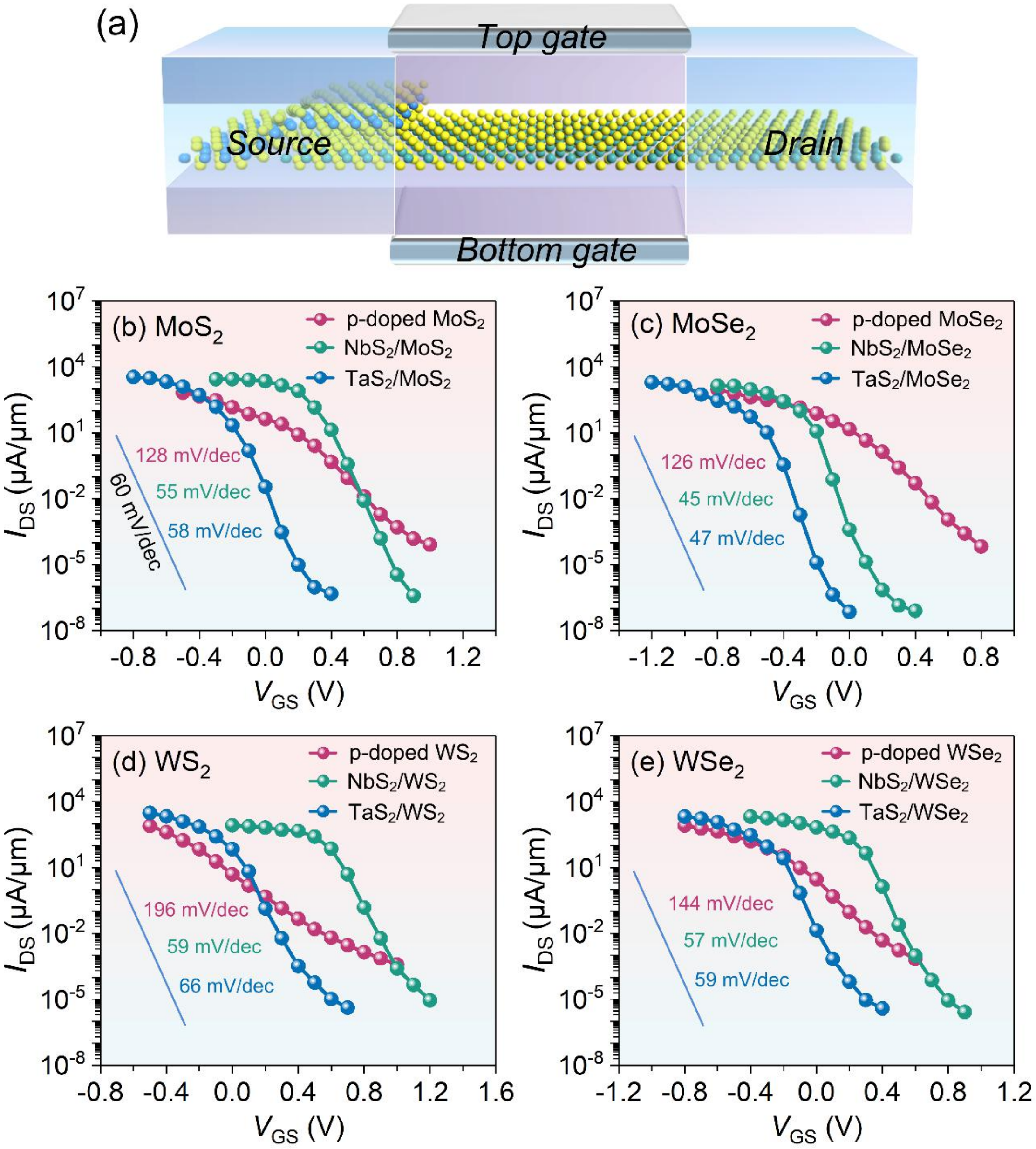

**Figure 2**. (a) The schematic view of top contact $MS_2$/TMDs CM-FET. (b-e) I-V curves of CM-FETs and p-doped TMD MOSFETs. SS of each I-V curve is labelled in (b-e). The blue lines labelled in (b-e) represent the SS of 60 mV/dec.

The schematic view of $MS_2$/TMDs CM-FET and the corresponding I-V curves are shown in Figure 2. The source and drain electrodes are modelled by “cold” metal and p-doped TMD materials, respectively, with the heterojunction of “cold” metal and TMD materials is shown in Figure S3. The p-type doping concentration of source is $3\times10^{13}$ $cm^{-2}$. Considering the top contact configuration in Figure 2a is more feasible than the edge contact configuration shown in Figure S4, we only calculate the performance of top contact CM-FETs in the next sections. I-V curves and performance benchmark of edge contact CM-FETs are shown in Figure S4 and Figure S5, respectively. The current of device is calculated by the Landauer–Bűttiker formula,[34]

$$I=\frac{2e}{h}\int_{-\infty}^{+\infty}\{T(E)[f_{S}(E-\varepsilon_{S})-f_{D}(E-\varepsilon_{D})]\}dE \tag{2}$$

where $T$(E) is the transmission coefficient, $f_S$ and $f_D$ are the Fermi-Dirac distribution functions for source and drain electrodes, $\varepsilon_S$ and $\varepsilon_D$ are the Fermi levels of source and drain electrodes, respectively. On-state current ($I_{on}$) is obtained by the corresponding on-state gate voltage ($V_{GS}$(on)). The $V_{GS}$(on) is defined as $V_{GS}$(on) = $V_{GS}$(off) + $V_{DS}$, where $V_{GS}$(off) is the off-state gate voltage defined by ITRS standards in 2013.[9] $V_{DS}$ is the bias voltage applied between the source and drain electrodes, fixed at 0.64 V in our work. It is observed that SS of CM-FETs significantly outperforms that of heavily p-doped TMD MOSFETs. Among these studied CM-FETs, $NbS_2$ CM-FETs exhibit lower SS because of the $E_{VG}$ close to the $\varepsilon_F$, while $TaS_2$ CM-FETs deliver larger currents owing to the wide transmission path between the $E_{VG}$ and $\varepsilon_F$ as shown in Figure 1b. The most favorable SS (45 mV/dec) and $I_{on}$ (2643 μA/μm) achieved by $NbS_2$/$MoSe_2$ and $TaS_2$/ $MoS_2$ CM-FETs respectively prove the above analysis. The top contact CM-FETs with smaller SS and $I_{on}$ than that of edge contact CM-FETs (data shown in Figure S4 and Figure S5) indicates that barriers of top contact devices interacted by the van der Waals force is higher than that of edge contact configurations. So, with the extreme low SS, the CMFET proposed in this work can fully sustain the more Moore roadmap in the 5nm nodes.

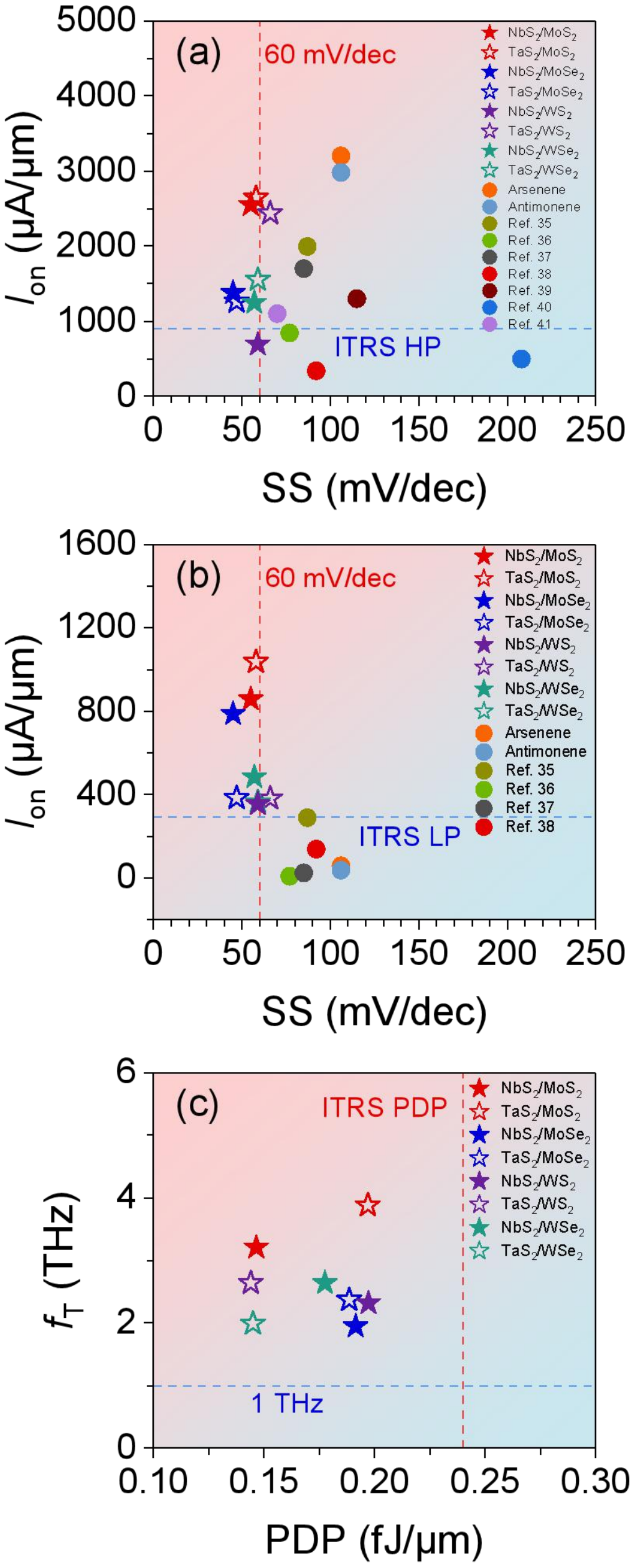

**Figure 3**. Benchmark of $I_{on}$ and SS of top contact CM-FETs for (a) HP goal and (b) LP goal, respectively. The other data are extracted from the arsenene,[35] antimonene,[35] BP,[36] Te,[37] GeSe,[38] silicane,[39] AsP,[40] Si nanowire (NW)[41] and carbon nanotube (CNT).[42] (c) The $f_T$ and PDP of CM-FETs compare with ITRS goals. The $I_{on}$ of ITRS HP and LP is set to 900 μA/μm and 295 μA/μm, respectively. The ITRS PDP is set as 0.24 fJ/μm.

To present a panoramic performance analysis of CM-FETs in 5 nm node, $I_{on}$ and SS of our work compared with previous results are shown in Figure3. The $I_{on}$ of CM-FETs ranges from 693 μA/μm to 2643 μA/μm and $TaS_2/MoS_2$ CM-FET with largest $I_{on}$ is about three times higher than that of ITRS HP goals. The $I_{on}$ performance is only inferior to the monolayer arsenene and antimonene MOSFETs ($I_{on}$ of 3200 μA/μm and 2980 μA/μm, respectively) and comparable with the black phosphorus (BP) MOSFET with 1994 μA/μm $I_{on}$. As is known, the BP is limited to the poor air-stability.[43] The superior $I_{on}$ of arsenene and antimonene MOSFETs benefits from the ideal heavily doped source/drain electrodes which can get rid of the influence of contact resistance.[35] As for SS, the performance of CM-FETs is overall superior to previous results in Figure 3. Especially, the $NbS_2/MoSe_2$ and $TaS_2/MoSe_2$ CM-FETs even achieve the low SS of 45 and 47 mV/dec, respectively at room temperature. In Figure 3c we plot the power dissipation (PDP) and cut-off frequency ($f_T$). PDP is a decisive parameter for low dissipation applications, defined as $\mathrm{PDP} = V_{\mathrm{DS}}(Q_{\mathrm{on}} - Q_{\mathrm{off}}) / w$ , where $Q_{on}$ and $Q_{off}$ are the total charge of the channel in on-state and off-state, respectively. $w$ is the channel width. The PDP values of CM-FETs vary from 0.144 to 0.197 fJ/um, obviously lower than the ITRS requirement of 0.24 fJ/um, showing a desired gate control capability. $f_T$ is a relevant factor for radio frequency devices, obtained by $f_{\mathrm{T}} = g_{\mathrm{m}} / (2\pi C_{\mathrm{G}})$ , where $g_m$ is the transconductance of MOSFETs, defined as $g_{\mathrm{m}} = I_{\mathrm{DS}} / V_{\mathrm{GS}}$ . $C_G$ is the gate capacitance, defined as $C_{\mathrm{G}} = \partial Q_{\mathrm{ch}} / \partial V_{\mathrm{GS}}$ , where $Q_{ch}$ is the charge in gate electrode region. All CM-FETs $f_T$ in Figure 3c reach the THz level, indicating that our CM-FETs are competent to be applied into radio frequency circuits.[44] Hence, the air-stability and excellent performance totally strengthen the competitive advantage of CM-FETs among these low-dimensional material MOSFETs as listed in Figure 3.

To better understand the mechanism of extreme low (sub-60 mV/dec) SS achieved by the CM-FETs, we take the $NbS_2/MoSe_2$ CM-FET as an example to plot the projected local density of states (PLDOS) in Figure 4. The SS ranges from 45 mV/dec to 59 mV/dec with $V_{GS}$ increasing form -0.2 V to 0 V in Figure 4a. The PLDOS in Figure 4b and 4c represents the switching process of the device. At $V_{GS}$ = -0.2 V, the $\Phi_B$ is small. Thermal emission current can directly transport through the channel region. With the increasing of $V_{GS}$, $\Phi_B$ and $E_{VG}$ are overlapped. Therefore, the current is only tunneling around the top of $E_{VG}$, so the transmission efficiency, $T$(E), around the $\varepsilon_S$ decreases rapidly compared with its counterpart in Figure 4b.

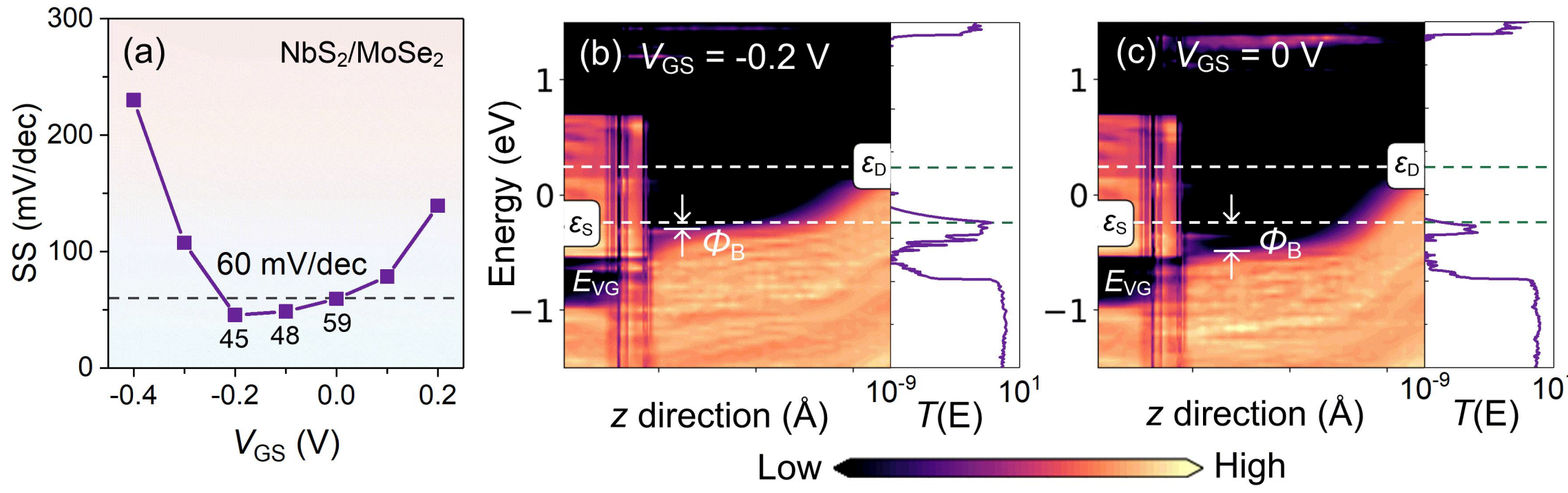

**Figure 4**. (a) The SS of $NbS_2/MoSe_2$ CM-FET as a function of $V_{GS}$. (b, c) Projected local density of states of $NbS_2/MoSe_2$ CM-FET with the barrier ($\Phi_B$) (b) above ($V_{GS}$ = -0.2 V) and (f) below ($V_{GS}$ = 0 V) the $E_{VG}$ of source.

## 2.2 The NDR effect of CM-FETs

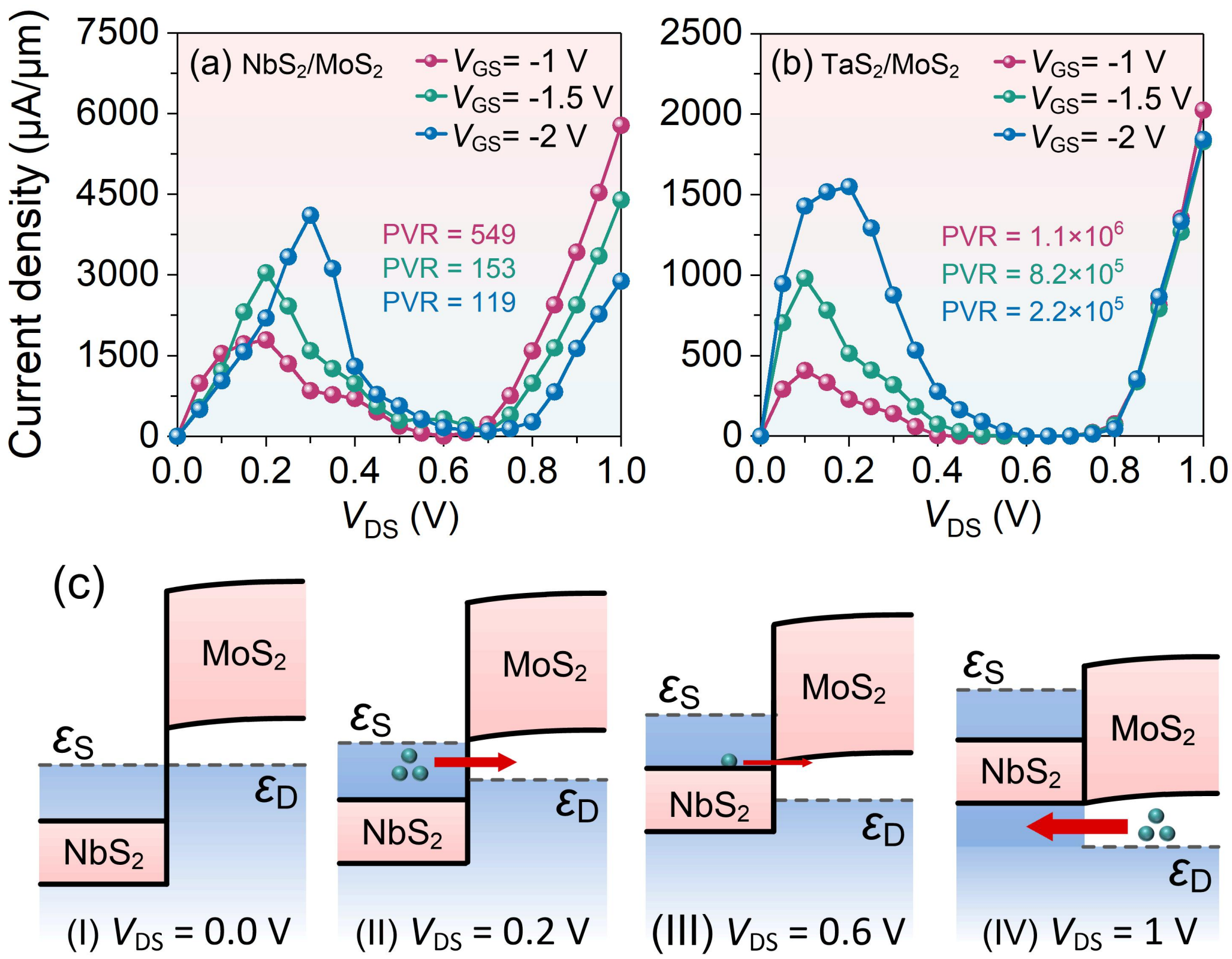

**Figure 5**. (a, b) The I-V curves of $NbS_2/MoS_2$ and $TaS_2/MoS_2$ CM-FETs. PVR labelled in (a, b) is the ratio of the peak current and valley current. (c) Energy band diagrams under different bias voltages ($V_{DS}$) at $V_{GS}$ = -1 V. The pink areas in (c) represent gaps of $NbS_2/MoS_2$ heterojunction. The $\varepsilon_S$ and $\varepsilon_D$ are the Fermi level of source and drain, respectively. The width of red arrows represents the magnitude of currents.

Apart from the superior $I_{on}$ and SS, CM-FETs are also capable to realize the NDR effect, which can be used in the analog circuits and is desirable for the multi-valued logic (MVL) system.[45] Based on the PLDOS in Figure 4, the “cold” metal source can form the type-III band alignment with the heavily p-doped $MoS_2$. I-V curves of $NbS_2/MoS_2$ and $TaS_2/MoS_2$ CM-FETs under various $V_{GS}$ are shown in Figure 5. Results claim that the $NbS_2/MoS_2$ CM-FET tends to deliver a large peak current from 1791 μA/μm to 4110 μA/μm, while the large peak-valley ratio (PVR) is readily to be obtained by the $TaS_2/MoS_2$ CM-FET with the detailed values labelled in Figure 5. The $NbS_2/MoS_2$

CM-FET fulfills the largest peak current 4110 μA/μm at $V_{GS}$ = -2V. The largest PVR of 1.1 × $10^6$ is achieved by the $TaS_2/MoS_2$ CM-FET at $V_{GS}$ = -1V, several orders of magnitude higher than the mainstream reports which will be discussed later. It is noteworthy that the $V_{GS}$ plays an important role in controlling the peak current and peak-valley ratio (PVR). The peak current varies inversely with the $V_{GS}$, while the PVR is proportional with the $V_{GS}$.

We plot the energy band diagrams at $V_{GS}$ = -1V in Figure 5c to analyze the NDR mechanism in the $NbS_2/MoS_2$ CM-FET. The I-IV in Figure 5c represent the energy band diagrams under various bias voltages ($V_{DS}$). Notably, unlike the conventional type-III band alignment NDR device with a narrow transmission path between the broken gap,[45-46] the wide transmission path, as shown in Figure S6, allows the CM-FET to achieve a large current, which is desirable for the extremely high PVR. Considering the $\varepsilon_S$ is already tuned to the valence band maximum (VBM) of $MoS_2$ with 0.2 V $V_{DS}$, the current reaches the peak point. With the increasing of $V_{DS}$, the current comes to the valley point, because the transmission is blocked by the overlapped gaps of $NbS_2$ and $MoS_2$. As the $V_{DS}$ further increases, the current starts to rise with a transmission path appearing below the VBM of $MoS_2$. It can be concluded that the width of energy window between the $\varepsilon_D$ and VBM of $MoS_2$ is a key factor for the peak current. Considering lower $V_{GS}$ corresponds to a wider transmission path as shown in Figure S7, the peak current is inversely proportional with the $V_{GS}$. In terms of the PVR, with the decreasing of $V_{GS}$, a larger $V_{DS}$ is needed to shift the gap of $MoS_2$ and reach valley point. The $\varepsilon_D$ in valley point is even close to the VBM of $NbS_2$, which leads to more efficient transmission through the VBM of $MoS_2$ and a larger valley current. Hence, the PVR is proportional with the $V_{GS}$.

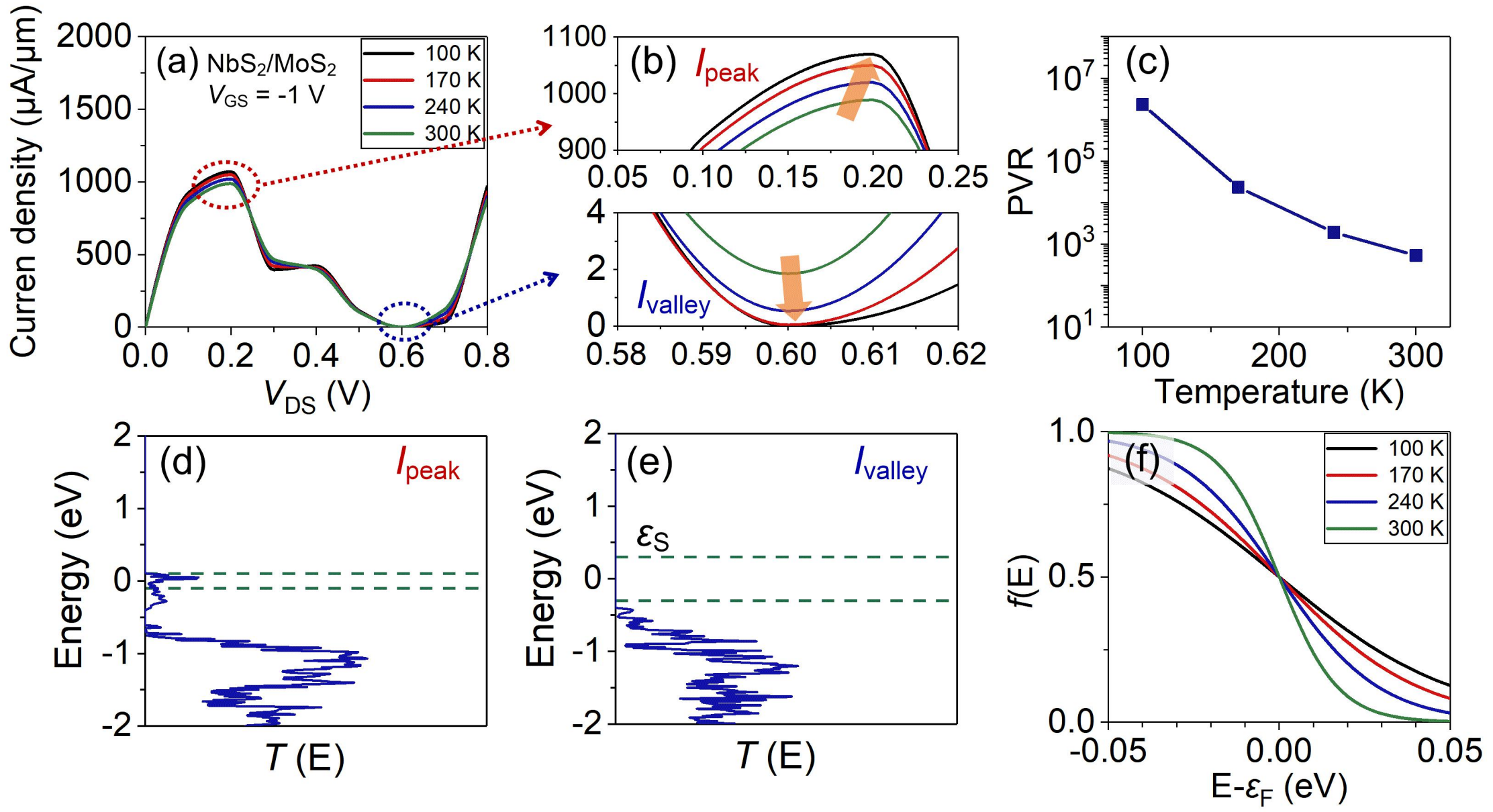

**Figure 6**. (a) I-V curves of the $NbS_2/MoS_2$ CM-FET at various temperature from 100 K to 300 K. (b) Enlarged views of I-V curves at peak point and valley point under different temperature. $I_{peak}$ and $I_{valley}$ represent the peak current and the valley current. (c) The PVR as a function of temperature from 100 K to 300 K. (d, e) The transmission spectrum of $NbS_2/MoS_2$ MOSFET at peak point and valley point, respectively. The $\varepsilon_S$ and $\varepsilon_D$ are the Fermi level of $NbS_2$ and $MoS_2$, respectively. The magnitude of $T$(E) indicates the transmission efficiency of carriers. (f) Fermi-Dirac distribution at different temperature.

Furthermore, to analyze the relationship between temperature and NDR effect, I-V curves of $NbS_2/MoS_2$ CM-FET with $V_{GS}$ = -1V at various temperature are shown in Figure 6. The corresponding analysis of $TaS_2/MoS_2$ CM-FET is shown in Figure S7. Encouragingly, the NDR effect is improved with the decreasing of temperature. As temperature decreases in Figure 6b, the peak current increases, while the valley current decreases. Consequently, the PVR(peak current) increases from 549(989 μA/μm) to 2.3 $\times 10^6$(1070 μA/μm) with temperature decreasing from 300 K to 100 K. Based on Eq. (2), the current is mainly contributed by transmission between the $\varepsilon_S$ and $\varepsilon_D$ of device. As temperature decreases in Figure 6f, the Fermi-Dirac distribution near the $\varepsilon_F$ becomes sharp. Thus, the weight of current contributed by the transmission between $\varepsilon_S$ and $\varepsilon_D$ is even higher, while the weight of transmission below the $\varepsilon_S$ and above the $\varepsilon_D$ becomes lower. In terms of peak current in Figure 6d, there is large $T$(E) around the $\varepsilon_S$. On the contrary, the $T$(E) of valley current between $\varepsilon_S$ and $\varepsilon_D$ is really flat. As a result, the valley current is proportional with the temperature and peak current varies inversely with the temperature. These results claim that the NDR performance of $NbS_2$/$MoS_2$ CM-FET is immune to the temperature oscillation, which enables the $NbS_2$/$MoS_2$ CM-FET adapt to more complex working environment.

Apart from the temperature effect, PVR and peak current are regarded as two decisive factors for practical applications. The NDR performance of $NbS_2$/$MoS_2$ and $TaS_2$/$MoS_2$ CM-FETs compared with previous results is summarized in Figure 7. To deliver a fair comparison, the device width is normalized to 10μm for both our devices and data from previous reports. The PVR of $NbS_2$/$MoS_2$ CM-FET is only inferior to the $MoS_2$/$WSe_2$ and BP/$Al_2O_3$/BP heterojunction NDR devices. However, the poor air-stability of BP and the small peak current of $MoS_2$/$WSe_2$ NDR device hinder their practical application. Compared with previous results superior in either PVR or peak current, the NDR devices in our work are fully capable to achieve both large PVR and peak current simultaneously.

Especially, the peak currents in this work are several orders of magnitude higher than previous results. The large peak current not only improves the noise margin ability of NDR devices and relevant circuits, but also increases the output power and enhances the stability of oscillation circuits. With the large peak current and PVR, $NbS_2/MoS_2$ and $TaS_2/MoS_2$ CM-FETs are suitable for the MVL system and multifunctional device in more than Moore field.

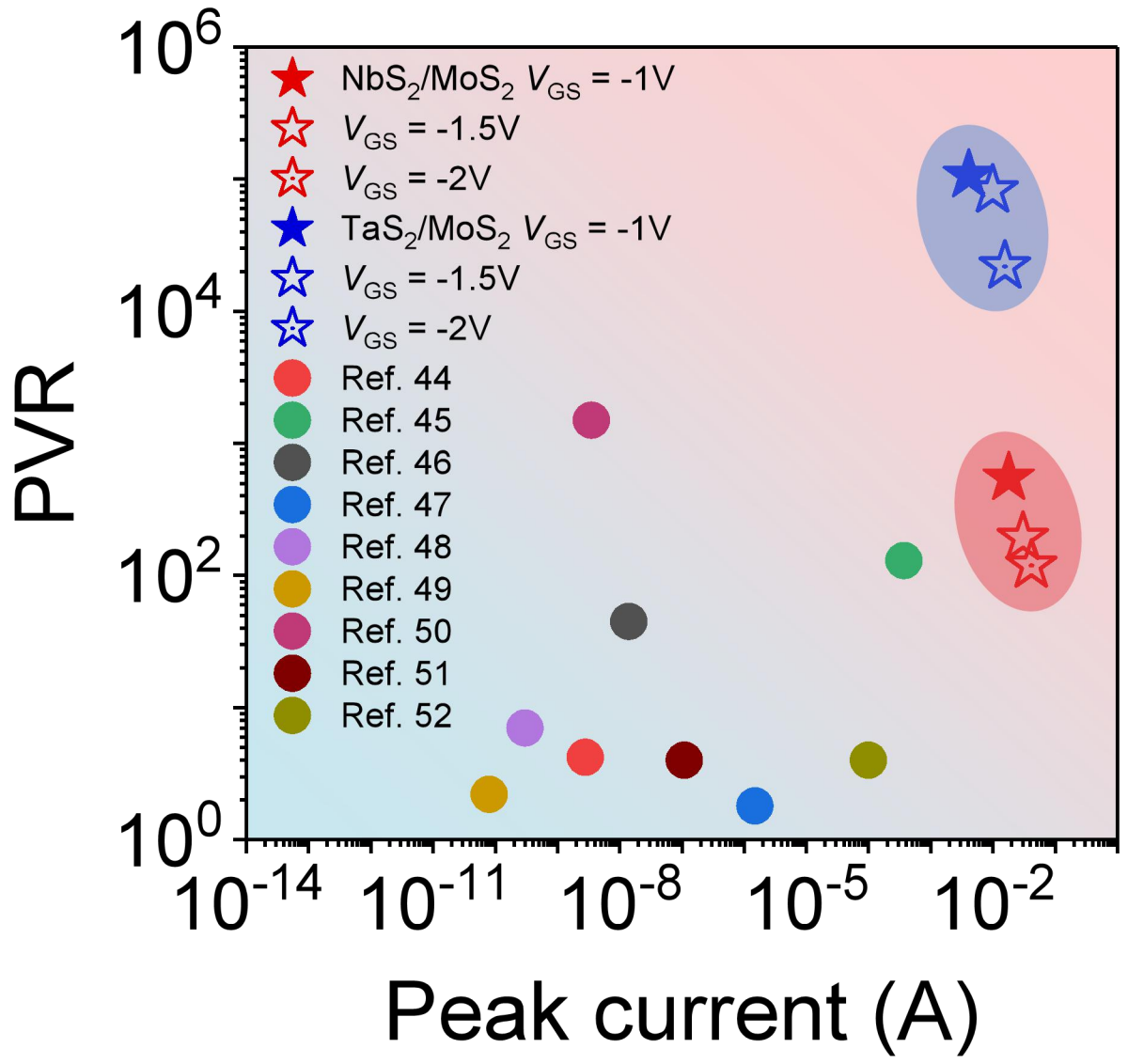

**Figure 7**. Benchmark of PVR as a function of peak current. Other data are extracted from the $BP/ReS_2$,[45] $MoS_2/WSe_2$,[46] Si/Ge,[47] $BP/SnSe_2$,[48] $BP/Al_2O_3/BP$,[49] Carbon quantum well,[50] $WSe_2/MoSe_2$,[51] Gra/BN/Gra[52] and $Gra/WSe_2/Gra$[53] NDR devices. The blue and red oval marks represent the performance variation of $NbS_2$ and $TaS_2$ CM-FETs, respectively.

## 3. CONCLUSIONS

In summary, we present a comprehensive electronic and transport calculation of nanoscale CM-FETs based on the "cold" metal $NbS_2$ and $TaS_2$ heterojunctions. The $I_{on}$ of CM-FETs with 5 nm channel length can achieve ITRS HP and LP goals simultaneously to further sustain the more Moore roadmap, which are rarely fulfilled in previous calculations because of the contradictory requirements of HP (high drain current) and LP (small SS). The dilemma is successfully solved by the superior switching performance of “cold” metals and the favorable mobility of 2D TMDs. The extreme low SS of 45 and 47 mV/dec is obtained for $NbS_2/MoSe_2$ and $TaS_2/MoSe_2$ CM-FETs, respectively In terms of NDR effect in more than Moore field, our results claim that the large peak current and PVR can be both achieved by $NbS_2/MoS_2$ and $TaS_2/MoS_2$ CM-FETs, owing to the wide broken gap feature, which is an obstacle of the Esaki diode. We find that the peak currents and PVR can be effectively controlled by the gate voltage and immune to the temperature influence. The recording 4110 μA/μm peak current and $1.1 \times 10^6$ PVR are achieved by $NbS_2/MoS_2$ and $TaS_2/MoS_2$ CM-FETs, respectively. The results prove that “cold” metal materials are competitive candidates for multifunctional logic devices and can be employed into the MVL system and radiofrequency circuits.

## 4. METHODOLOGY

We calculated the electronic and transport properties of “cold” metal $MS_2$ and their devices with TMD channels based on density functional theory (DFT) and non-equilibrium Green function (NEGF) with Atomsitix Tool Kit 2020 package.[54] The exchange correlation was the Perdew-Burke-Ernzerhof (PBE) functional of generalized gradient approximation (GGA). All simulations are conducted with the Pseudo Dojo pseudopotential and DFT-D3 van der Waals correction.[55] A 80-Hartree cut-off energy was adopted. Monkhorst-Pack grids used for the transport calculation sampling the 8×1×163 k point meshes. To avoid the interaction from adjacent layers, a 30 Å vacuum was applied to the device for transport calculations. The other basic settings refer to our previous work.[56-58]

## AUTHOR INFORMATION

### Corresponding Author

* E-mail: yguo@whu.edu.cn

* E-mail: zz389@cam.ac.uk

### Supporting Information

Supporting Information is available from the Wiley Online Library or from the author.

## ACKNOWLEDGMENTS

This work is supported by Wuhan University. Authors acknowledge the financial support from the Fundamental Research Funds for the Central Universities, and the funding from EPSRC grant EP/P005152/1. The numerical calculations in this work is conducted on the supercomputing system in the Supercomputing Center of Wuhan University.

### Conflict of Interest

The authors declare no conflict of interest

# TOC

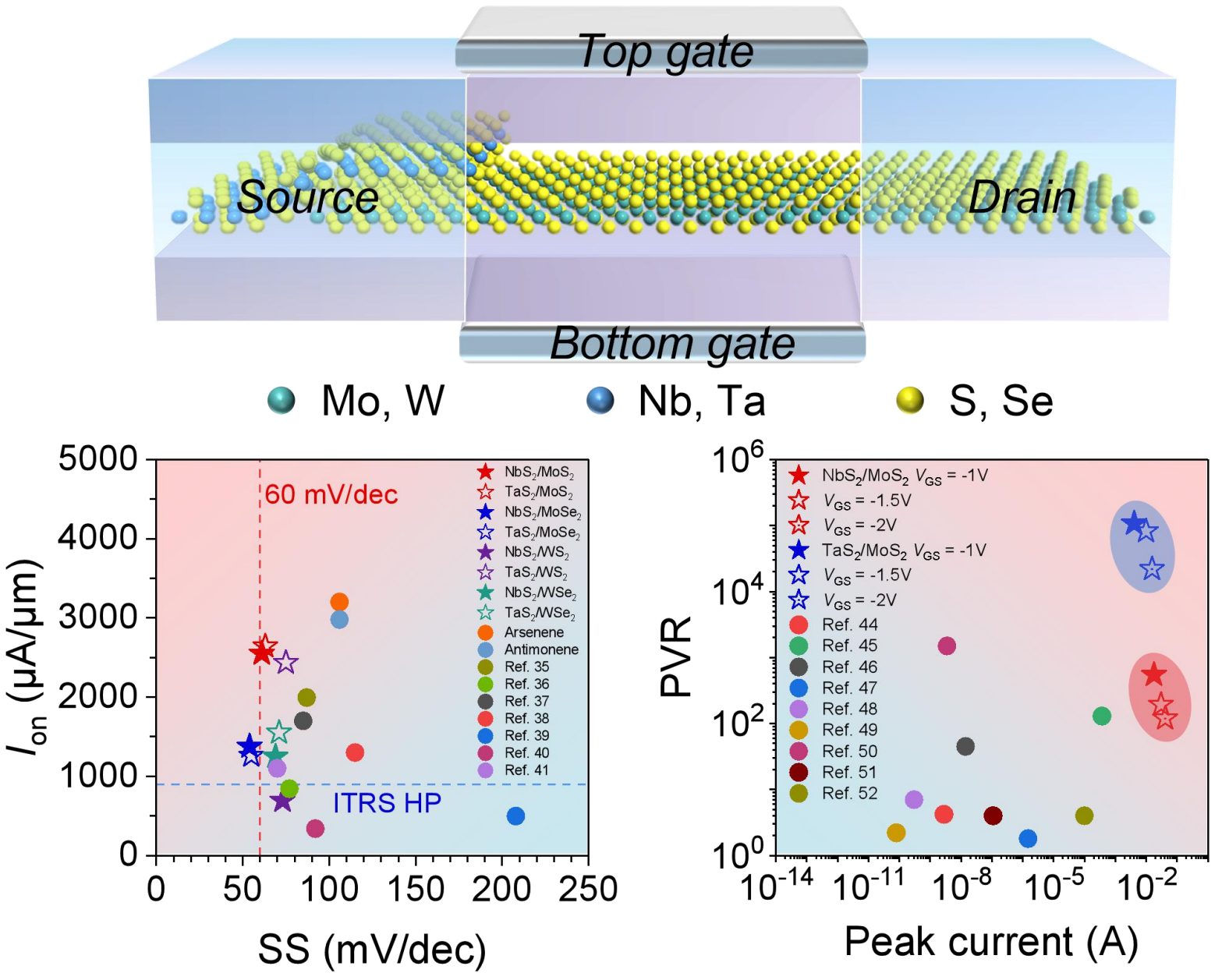